# Thermal expansion of the magnetorefrigerant $Gd_5(Si,Ge)_4$

M. Nazih, A. de Visser*, L. Zhang, O. Tegus and E. Brück

*Van der Waals-Zeeman Institute, University of Amsterdam,
Valckenierstraat 65, 1018 XE Amsterdam, The Netherlands*

**Abstract**

We report thermal expansion measurements carried out on a single-crystal of the giant magnetocaloric effect material $Gd_5(Si_{0.43}Ge_{0.57})_4$. At the magneto-structural phase transition at $T_0 \sim 240$ K, large steps in the relative length change $\Delta L/L$ along the principle crystallographic axes are observed. The single-step behaviour in $\Delta L/L$ indicates that the magnetic and structural phase transition occur at one and the same temperature. The specific heat was measured in order to determine the latent heat. Using the Clausius-Clapeyron relation we extract a hydrostatic pressure dependence $dT_0/dp$ of $3.2 \pm 0.2$ K/kbar. The pressure effect is strongly anisotropic as follows from the anisotropy in the linear thermal expansion. Uniaxial pressure along the a-axis enhances $T_0$, while uniaxial pressure in the *bc*-plane suppresses $T_0$.

*Corresponding author:
Phone: +31-20-5255732.
Fax: +31-20-5255788.
E-mail addresss: devisser@science.uva.nl.

# 1. Introduction

In 1997 Pecharsky and Gschneidner [1] reported the discovery of a giant record-high magnetocaloric effect in the intermetallic series $Gd_5(Si_xGe_{1-x})_4$. The observation that for $0.4 \leq x \leq 0.5$ the magnetocaloric effect (MCE) takes place in the sub-room temperature range $T \sim 240\text{-}290$ K is of significant technological relevance as it identifies these compounds as potential candidates for the application as magnetorefrigerant. Nowadays, there is much interest in magnetic refrigeration, as it offers the prospect of an energy efficient and environment friendly alternative for the traditional vapour-cycle refrigeration technique. The development of magnetic cooling near room temperature requires basic research into the exploration and characterisation of new magnetic refrigerants, as well as into the design of new magnetic-refrigeration devices [2]. Recently, a new promising material was discovered [3], namely the transition-metal compound MnFe(P,As), which exhibits a giant MCE just above room temperature ($T \sim 300$ K).

In this paper we concentrate on the $Gd_5(Ge_{1-x}Si_x)_4$ system. The magnetic and room-temperature structural properties of $Gd_5(Ge_{1-x}Si_x)_4$ compounds have been reported in Ref.4. $Gd_5(Ge_{1-x}Si_x)_4$ compounds crystallize in the orthorhombic structure (space group Pnma), except for the composition range $0.25 \leq x \leq 0.5$ where the room-temperature crystal structure is monoclinic (space group P112$_1$/a). This is the composition range of interest, as here the giant MCE is observed. The parent compound $Gd_5Si_4$ shows ferromagnetic order with a Curie temperature $T_C = 335$ K, while $Gd_5Ge_4$ is a ferrimagnet with an ordering temperature of 125 K. By diluting $Gd_5Si_4$ with Ge, $T_C$ decreases slowly to a value of ~276 K at $x = 0.5$. Upon further alloying $T_C$ decreases more rapidly at a rate of ~5 K/at.%Ge till $x = 0.25$.

The giant MCE was discovered in a polycrystalline sample with composition $x = 0.5$ [1] and initially attributed to two consecutive ferromagnetic phases transitions with Curie temperatures of 299 K and 276 K. However, it soon became clear that the first transition was due to an impurity phase [5]. By conducting temperature dependent x-ray diffraction experiments on a polycrystalline sample with $x = 0.45$, Morellon and co-workers [6] discovered that the magnetic phase transition ($T_C = 240$ K) is accompanied by a first-order structural transition, which restores the orthorhombic structure of the parent compounds. Hence, the giant MCE is related to a magneto-structural transition. The structural transition is unusual in the sense that the low temperature phase has a higher symmetry (orthorhombic) than the high-temperature phase (monoclinic). A detailed characterization of the first-order magneto-structural transition is of relevance for the understanding of the MCE in the $Gd_5(Ge_{1-x}Si_x)_4$ alloys.

In this paper we present the first study of the thermal properties of single crystalline $Gd_5(Ge_{1-x}Si_x)_4$. We focus on the thermal expansion of a single-crystalline sample with composition $Gd_5(Si_{0.43}Ge_{0.57})_4$ in the temperature range 5-290 K. We compare our results with specific heat and magnetisation data measured on the same single crystal. Thermal expansion data of a polycrystalline sample ($x = 0.45$) [6] showed a fairly broad transition ($\Delta T \approx 15$ K), with some structure at the low temperature side, and a volume contraction $\Delta V/V$ upon cooling of ~0.4%. The main objectives of our experiments are (i) to investigate the anisotropy in the thermal expansion coefficients, and (ii) to investigate whether the magnetic and structural transitions coincide and form one single magneto-structural transition.

## 2. Experimental

A single-crystalline sample was prepared in a two-mirror furnace (NEC) under argon atmosphere by means of the travelling floating zone method. As starting materials we used Gd with purity 4N and Ge and Si with purity 6N. The grown single-crystalline rod had a diameter of 4 mm and a length of 20 mm. Electron probe microanalysis showed a small variation of the composition along the rod with an average composition $Gd_5(Si_{0.43}Ge_{0.57})_4$. The powder diffraction pattern could be indexed within the monoclinic structure (spacegroup $P112_1/a$) with lattice parameters $a = 7.585$ Å, $b = 14.800$ Å and $c = 7.777$ Å and a monoclinic angle $\beta = 93.29°$. The unit cell contains 36 atoms (4 formula units) and has a volume $\Omega = 871.6$ Å$^3$. The molar volume $V_m = 1.312 \times 10^{-4}$ m$^3$/mol. These crystallographic data are in good agreement with literature values [4].

For the thermal expansion measurements a rectangular bar, with plane-parallel surfaces perpendicular to the main crystallographic axes, was cut from the central part of the as-grown crystalline rod by means of spark-erosion. Since the monoclinic distortion is small, we retain orthorhombic labelling of the axes. The dimensions of the sample amount to $a \times b \times c = 5.0 \times 3.4 \times 2.8$ mm$^3$. Specific heat measurements were performed on the same sample, while magnetization data were taken on a smaller piece cut from the same sample.

The thermal expansion $\Delta L/L$ along the $a$, $b$ and $c$ axis was measured in the temperature range 5-290 K using a parallel plate capacitance dilatometer machined of copper [7]. The data were corrected for the cell effect, i.e. the signal of the cell with a dummy copper sample mounted. The measurements were performed by slowly heating or cooling at a typical rate of 0.2 K per minute. Magnetisation measurements as a function of temperature were carried out using a commercial squid magnetometer (Quantum Design). The specific heat was measured by a semi-adiabatic technique in a home-built calorimeter.

## 3. Results

In order to characterize the ferromagnetic transition of our single crystalline sample we measured the temperature variation of the magnetization $M_i(T)$ in small magnetic fields applied along the $i = a$, $b$ and $c$-axis. In Fig.1 we present the data taken with increasing temperature in a magnetic field $B = 0.05$ T. The sharp increase of $M(T)$ signals the Curie temperature $T_C = 240.4 \pm 1.0$ K. The sharp step-wise change of the magnetization is a clear indication of the first-order nature of the transition. The apparent low-field anisotropy in $M(T)$ in the ordered state is mainly due to different demagnetizing factors for the different directions. In a field of 5 T $M(T)$ is close to isotropic [4,8]. The peculiar increase of $M_a(T)$ in the ordered state possibly indicates a spin-reorientation process. In the paramagnetic state $M(T)$ obeys the Curie-Weiss law with an effective moment $\mu_{eff}$ of 8.2±0.1, 8.6±0.1 and 8.4±0.1 $\mu_B$/Gd-atom for the $a$, $b$ and $c$ direction. The extracted effective moment is slightly larger than the theoretical value 7.94 $\mu_B$/Gd-atom. The paramagnetic Curie temperature $\theta_p$ amounts to 204±1 K. Notice that these magnetic parameters are determined for the monoclinic structure, while the ferromagnetic phase adopts the orthorhombic structure. The magnetization data presented in Fig.1 are in excellent agreement with the results reported in Refs.5 and 7 for samples with slightly different Ge/Si ratios.

The main results of the thermal expansion measurements are presented in Figs.2 and 3. In Fig.2 we show the relative thermal expansion $\Delta L/L \equiv (L(T)-L_0)/L_0$, where $L_0 = L(5\ K)$, along the $a$, $b$ and $c$-axis. The magneto-structural transition is characterized by large step-wise changes in $\Delta L/L$ for all crystallographic directions at a temperature $T_0 = 240.0 \pm 1.0$ K. The thermal expansion shows a pronounced anisotropy between the $bc$-plane and the $a$-axis behaviour. At $T_0$, upon heating, the step in $\Delta L/L$ for the $b$ and $c$-axis attains negative values of -2.0x10$^{-3}$ and -2.1x10$^{-3}$, respectively, while for the $a$-axis the step is positive and much larger ~6.8x10$^{-3}$. Consequently, the volume change $\Delta V/V = (\Delta L/L)_a + (\Delta L/L)_b + (\Delta L/L)_c$ is positive and amounts to 2.7x10$^{-3}$ (see inset in Fig.2). In Fig.3 we compare the thermal expansion data taken with increasing and decreasing temperatures. The data are identical, except near the magneto-structural transition. Upon lowering the temperature, the magneto-structural transition takes place at $T_0 = 236.0 \pm 1.0$ K. The hysteresis associated with the first-order transition amounts to $\Delta T = 4.0$ K.

In Fig.4 we show the specific heat at constant pressure $c_p$ (per mol formula unit) measured with increasing temperature. The magneto-structural transition takes place at $T_0 = 239.7 \pm 1.0$ K. The entropy change $\Delta S$ at the transition is $11.0 \pm 0.5$ J/molK and the latent heat $L = 2.63 \pm 0.12$ kJ/mol.

## 4. Analysis and discussion

A close inspection of the thermal expansion data in the temperature region near 240 K shows that the phase transformation occurs as a step-wise change in $\Delta L/L$, without additional structure, and at identical temperatures for the *a*, *b* and *c*-axis. The sharp step-wise changes confirm the first-order character of the phase transition. Since the structural transition temperature $T_0 = 240.0 \pm 1.0$ K coincides with the Curie temperature $T_C = 240.4 \pm 1.0$ K, within the experimental error, we conclude that the magnetic and structural transitions take place at the same temperature. Our results confirm the conclusions reached in Ref.6 based on experiments on polycrystalline samples, namely the transformation takes place at one single temperature $T_0 = 240$ K and is from a high-temperature monoclinic paramagnetic phase to a low-temperature orthorhombic ferromagnetic phase. The tight link between the magnetic and structural transition is further supported by thermal expansion experiments in an external magnetic field, which showed that the volume change $\Delta V/V$ shifts up in temperature at a rate 4.5 K/T for a polycrystalline sample with $x = 0.45$ [6].

The large anisotropy in the steps in $\Delta L/L$ at $T_0$ is in agreement with the variation of the lattice parameters with temperature as determined by x-ray diffraction [4,9]. Choe and co-workers (Ref.9) investigated the structural details of the transformation by x-ray diffraction experiments on a single crystal with $x = 0.50$. They reported that the orthorhombic to monoclinic distortion occurs by a shear mechanism in which the covalent bonds between Si and Ge atoms that are richer in Ge increase their distances by 0.859 Å. In fact, half of these dimers in the orthorhombic phase are broken in the monoclinic phase. This shear movement takes place along the *a*-direction, which explains the large increase in $\Delta L_a/L_a$. Consequently, the *b*- and *c*-directions show negative and smaller steps in $\Delta L/L$. The resulting volume change $\Delta V/V = 2.7 \times 10^{-3}$ is somewhat smaller than the value $\sim 4 \times 10^{-3}$ reported for a polycrystalline sample with $x = 0.45$ [6].

Next we evaluate the pressure dependence of $T_0$ with help of the Clausius-Clapeyron relation $dT_0/dP = \Delta V/\Delta S$. With the experimental values $\Delta V = 3.4(\pm 0.1) \times 10^{-3}$ m$^3$/mol and $\Delta S = 11.0 \pm 0.5$ J/molK we obtain $dT_0/dP = 3.2 \pm 0.2$ K/kbar. This value is in good agreement with the value $dT_0/dP = 3.46$ K/kbar extracted from thermal expansion measurements under hydrostatic pressure for a sample with $x = 0.45$ [6]. A positive value of $dT_0/dP$ could already be predicted by comparing the unit cell volume $\Omega$ and $T_0$ for the Gd$_5$(Si$_x$Ge$_{1-x}$)$_4$ compounds in the concentration range $0.24 < x < 0.5$ [4]. By using the data of Ref.4 we obtain $V^{-1}dT_0/dV = -8.7 \times 10^{-5}$ K. Combining the latter value with the experimental value of $dT_0/dP$, we calculate a compressibility $\kappa = -V^{-1}dV/dp = 0.3 \times 10^{-11}$ m$^2$/N. One should bear in mind however, that the pressure effects are strongly anisotropic. Uniaxial pressure along the a-axis will result in an increase of $T_0$, while uniaxial pressure in the $bc$-plane will result in a decrease of $T_0$. This provides important directions how to chemically substitute the system in order to further enhance $T_0$.

The specific heat shown in Fig.4 is dominated by the phonon contribution. Till about 70 K the specific heat can be described adequately by the Debye function with a Debye temperature $\theta_D = 237$ K, as illustrated by the solid line in Fig.4, which represents the sum of the Debye contribution and a small linear electronic term ($\gamma T$) with a Sommerfeld coefficient $\gamma = 0.03$ J/molK$^2$. In the high-temperature range ($T > T_0$), the Debye function saturates at the value $3rR = 225$ J/molK, where $R$ is the gas constant and $r = 9$ is the number of atoms per formula unit. The measured specific heat, after correcting for the electronic term, amounts to 235 J/molK at $T = 250$ K, which is in good agreement with the model value $3rR$. After subtracting the phonon and linear electronic contribution from the measured specific heat the "magnetic" specific heat, $c_m$, is obtained. This contribution becomes visible for temperatures $T > 70$ K and is attributed to the magnetic ordering of the Gd 4$f$ moments and the structural transition at $T_0$. The total magnetic entropy obtained by integrating the $c_m/T$ versus $T$ data up to 245 K equals 41 J/molK. Notice that this value is considerably smaller than the maximum magnetic entropy for the Gd$_5$(Si$_x$Ge$_{1-x}$)$_4$ compounds, taking into account the ($2J+1$) degeneracy of the ground state multiplet with angular momentum $J = 7/2$, amounts to $5R\ln(2J+1) = 86$ J/molK.

The thermal expansion and the specific heat data can be compared by calculating the temperature dependent effective Grüneisen parameter $\Gamma_{eff} = \beta(T)V_m/\kappa c(T)$. Here $\beta(T) = \Sigma_i \alpha_i(T)$ is the coefficient of volume expansion and $\alpha_i = L_i^{-1} dL_i/dT$ is the coefficient of linear thermal expansion along the different crystallographic axes. $\Gamma_{\text{eff}}$ shows a weak increase with temperature and ranges from 4.2 up to 5.6 in the temperature range 20-200 K (where we used $\kappa = 0.3 \times 10^{-11}$ m$^2$/N). This is about a factor 2-3 larger than the value $\Gamma_{\text{ph}} = 2$ expected for a simple metal with a dominant phonon contribution.

## 5. Summary

We have measured the thermal expansion of a single-crystalline sample of the magneto-refrigerant $Gd_5(Si_xGe_{1-x})_4$ with $x = 0.43$. At the magneto-structural phase transition at $T_0 = 240.0 \pm 1.0$ K large steps in the relative length change $\Delta L/L$ along the principle crystallographic axes are observed. The thermal expansion is strongly anisotropic. The step in $\Delta L_a/L_a$ measured with increasing temperature is positive, while the steps in $\Delta L_b/L_b$ and $\Delta L_c/L_c$ are negative. The resulting step in the volume change $\Delta V/V$ at $T_0$ is positive and amounts to $\sim 2.7 \times 10^{-3}$. The thermal expansion in the vicinity of $T_0$ shows sharp steps in $\Delta L/L$ and no additional structure, which indicates that the magnetic and structural phase transition occur at one and the same temperature. By combining specific heat and thermal expansion data and by making use of the Clausius-Clapeyron relation we extract a hydrostatic pressure dependence $dT_0/dp$ of $3.2 \pm 0.2$ K/kbar. The pressure effect is strongly anisotropic: uniaxial pressure along the a-axis enhances $T_0$, while uniaxial pressure in the *bc*-plane suppresses $T_0$.

**Figure captions**

Fig. 1  Magnetisation as a function of temperature of single-crystalline $Gd_5(Si_{0.43}Ge_{0.57})_4$ in an external field $B = 0.05$ T directed along the crystallographic axes as indicated.

Fig. 2  Linear thermal expansion of $Gd_5(Si_{0.43}Ge_{0.57})_4$ plotted as $\Delta L/L$ along the $a$, $b$ and $c$-axis as indicated. The inset in the top panel shows the volume thermal expansion.

Fig. 3  Linear thermal expansion of $Gd_5(Si_{0.43}Ge_{0.57})_4$ along the $a$, $b$ and $c$-axis as indicated. Data are shown for increasing (closed circles) and decreasing (open circles) temperatures. The thermal hysteresis amounts to 4 K.

Fig. 4  Specific heat at constant pressure of single-crystalline $Gd_5(Si_{0.43}Ge_{0.57})_4$. The solid line represent the sum of a Debye function with $\theta_D = 237$ K and a linear electronic term with $\gamma = 0.03$ J/molK$^2$. The inset shows the peak in $c_p$ at the magnetostructural transition.

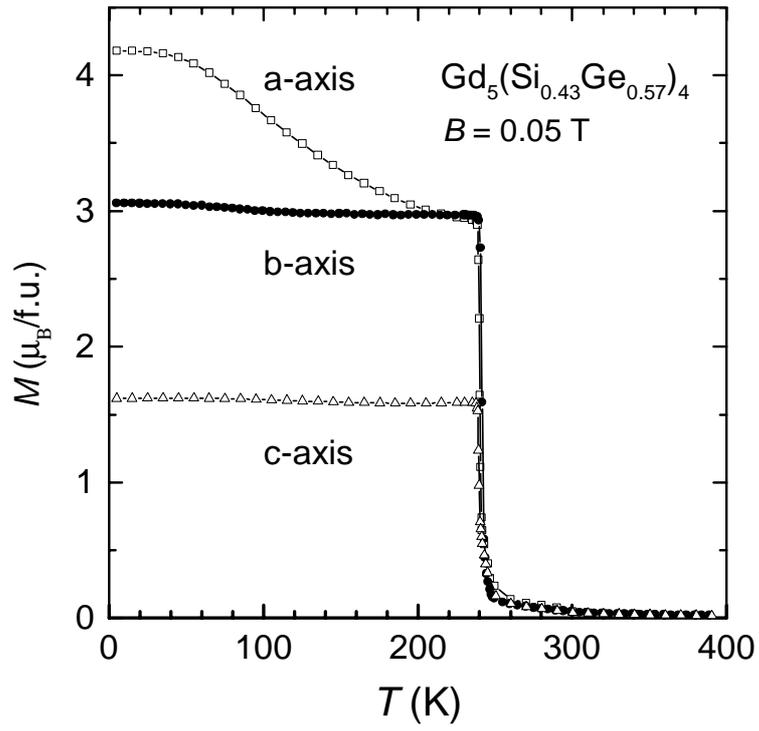

Figure 1

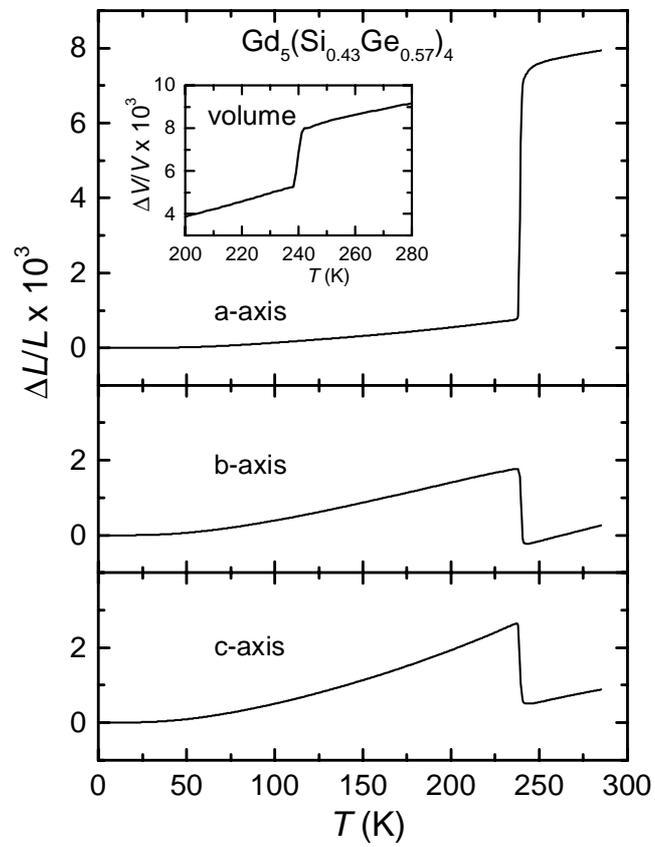

Figure 2

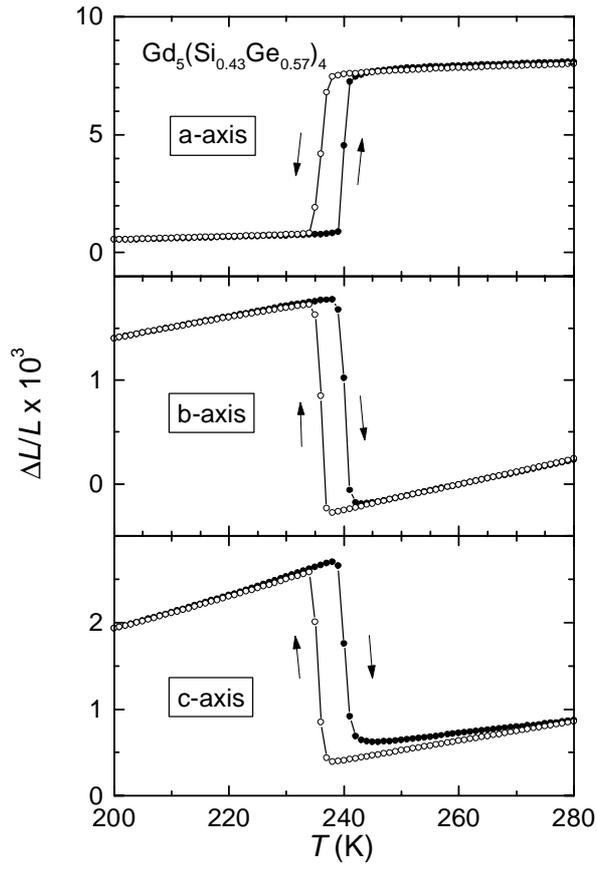

**Figure 3**

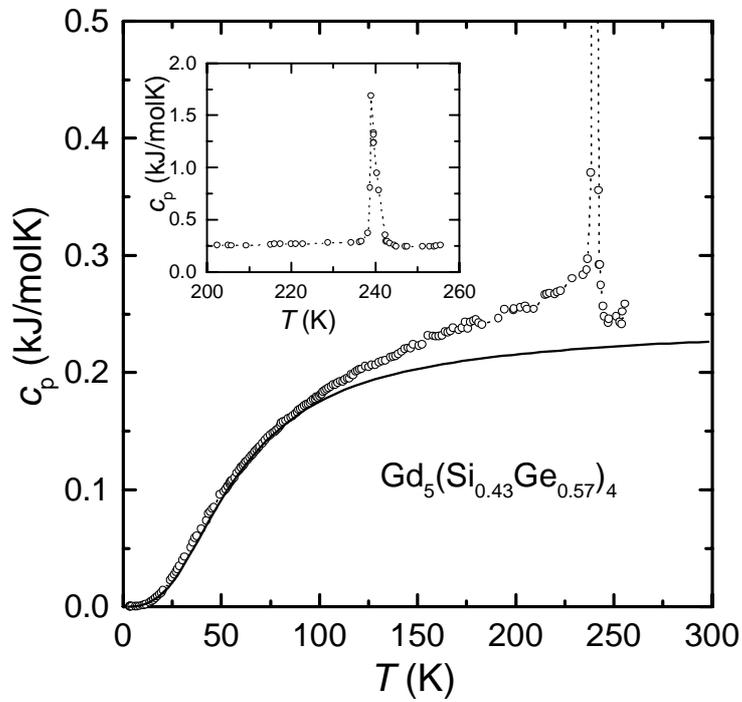

**Figure 4**